%% file: masterfile.tex
\documentclass[10pt,twoside,BCOR7mm,DIV15,headinclude,footexclude,cleardoubleempty,idxtotoc]{scrartcl}

\usepackage[english]{babel}
\usepackage{graphicx}
\usepackage{hyperref}
\usepackage{scrpage2}
\usepackage{hyperref}
\usepackage{ifthen}

\makeatletter
\renewcommand{\@biblabel}[1]{}
\renewcommand{\@cite}[2]{%
{#1\ifthenelse{\boolean{@tempswa}}{,#2}{}}}
\makeatother

\hypersetup{breaklinks=true
,colorlinks=true,linkcolor=black,urlcolor=blue
,citecolor=black}

\ofoot{\thepage}
\ifoot{}

\setheadsepline{1pt}

\setkomafont{pagehead}{\normalfont\sffamily}
\setkomafont{pagenumber}{\normalfont\rmfamily}

\usepackage{booktabs}
\usepackage{amsmath}
\usepackage{amssymb}
\usepackage{multicol}
\usepackage{float}

\makeatletter
\newcommand{\listofcontributions}{\@starttoc{con}}

\newcommand{\l@contribution} {\@dottedtocline{1}{1.5em}{2.3em}}
\makeatother

\newenvironment{contribution}{
\setcounter{section}{0}
\setcounter{figure}{0}
\setcounter{table}{0}
\begin{flushleft}
{\em Clumping in Hot Star Winds \\
W.-R.\ Hamann, A.\ Feldmeier \& L.\ Oskinova, eds.\\
Potsdam: Univ.-Verl., 2007 \\
URN: http://nbn-resolving.de/urn:nbn:de:kobv:517-opus-13981
} 
\end{flushleft}
}{
\newpage
\lehead{}
\rohead{}
}

\begin{document}

\setlength{\baselineskip}{2.5ex}

\begin{contribution}
\input{myarticle.tex}
\end{contribution}


\end{document}

%% file: myarticle.tex

\newcommand\ionalt[2]{#1$\;${\small\rmfamily #2}\relax}

\lehead{M.\ A.\ Leutenegger, D.\ H.\ Cohen, S.\ M.\ Kahn, S.\ P.\ Owocki, F.\
  B.\ S.\ Paerels}

\rohead{Resonance scattering in X-ray line profiles of $\zeta$ Pup}

\begin{center}
{\LARGE \bf Resonance scattering in the X-ray emission lines profiles of $\zeta$ Puppis}\\
\medskip

{\it\bf M. A. Leutenegger$^1$, D. H. Cohen$^2$, S. M. Kahn$^3$,
  S. P. Owocki$^4$, \& F. B. S. Paerels$^1$}\\

{\it $^1$Columbia Astrophysics Laboratory, USA}\\
{\it $^2$Swarthmore College, USA}\\
{\it $^3$Stanford University and Stanford Linear Accelerator Center, USA} \\
{\it $^4$Bartol Research Institute, USA}

\begin{abstract}
  We present XMM-{\it Newton} Reflection Grating Spectrometer observations of
  pairs of X-ray emission line profiles from the O star $\zeta$ Pup
  that originate from the same He-like ion. The two profiles in each pair have
  different shapes and cannot both be consistently fit by models assuming the
  same wind parameters. We show that the differences in profile shape can be
  accounted for in a model including the effects of resonance scattering,
  which affects the resonance line in the pair but not the intercombination
  line. This implies that resonance scattering is also important in single
  resonance lines, where its effect is difficult to distinguish from a low
  effective continuum optical depth in the wind. Thus, resonance scattering
  may help reconcile X-ray line profile shapes with literature mass-loss
  rates. 
\end{abstract}
\end{center}

\begin{multicols}{2}

\section{Introduction}

Recent studies of X-ray emission line Doppler profiles from O stars have found
that the profiles are too symmetric to be explained in the context of a
smooth-wind model, assuming the published mass-loss rates of O stars are
correct (e.g. Kramer et al. \cite{KCO03}, Cohen et al. \cite{Cohen06}).

There is mounting evidence from studies of UV absorption profiles as well as
other lines of inquiry that the true mass-loss rates of O stars are at least a
factor of a few smaller than those derived from density-squared diagnostics
(e.g. Bouret et al. \cite{BLH05}, Fullerton et al. \cite{FMP06}). However,
some of the observed X-ray profiles appear to require mass-loss rate
reductions of an order of magnitude (e.g. Kramer et al. \cite{KCO03}, Cohen et
al. \cite{Cohen06}). 

Two alternative explanations for X-ray profile shapes may mitigate the
requirement for large mass-loss rate reductions. {\it Porosity}, the formation
of very large, optically thick clumps, could reduce the effective opacity of
the wind to X-rays (e.g. Oskinova et al. \cite{OFH06}, Owocki \& Cohen
\cite{OC06}). {\it Resonance Scattering} could change the local angular
distribution of emitted X-rays, symmetrizing line profiles and mimicking the
symmetric profiles of a wind with little absorption (Ignace \& Gayley
\cite{IG02}).

\section{He-like emisison line profile discrepancy}

\begin{figure}[H]
  \begin{center}
    \includegraphics[width=\columnwidth]{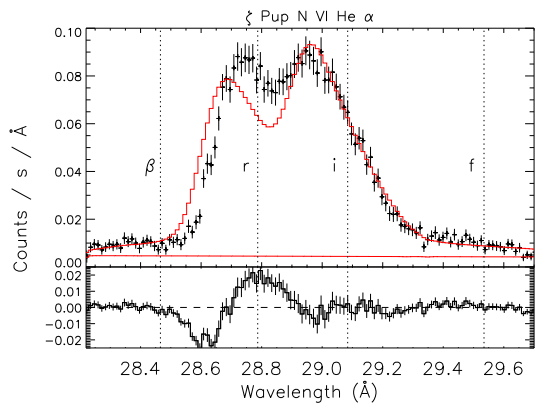}
    \caption{\ionalt{N}{VI} He-like triplet from $\zeta$~Pup observed with RGS.
      The model (red line) does not include resonance scattering and is chosen
      to fit the red wing of the intercombination line. The dashed lines
      labelled $r$, $i$, $f$, and $\beta$ indicate the rest wavelengths of the
      resonance, intercombination, and forbidden lines, and the \ionalt{C}{VI}
      Ly$\beta$ line, respectively. The lower red line gives the continuum
      strength. } 
    \label{fig:nvi_nors}
  \end{center}
\end{figure}

\begin{figure}[H]
  \begin{center}
    \includegraphics[width=\columnwidth]{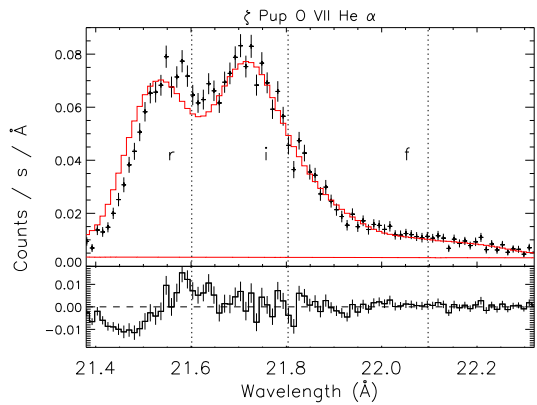}
    \caption{Same as Fig.~\ref{fig:nvi_nors}, but for \ionalt{O}{VII}.}
    \label{fig:ovii_nors}
  \end{center}
\end{figure}

In Leutenegger et al. (\cite{Leut07}) we present over 400 ks of net
XMM-{\it Newton} Reflection Grating Spectrometer (RGS) data from observations
of $\zeta$ Pup. This is by far the highest signal-to-noise high-resolution
X-ray spectrum available for any O star. In figures~\ref{fig:nvi_nors} and
\ref{fig:ovii_nors} we show this data with the line profile models for the
\ionalt{N}{VI} and \ionalt{O}{VII} He-like triplets of $\zeta$ Pup,
respectively. The profile model is described in Owocki \& Cohen (\cite{OC01})
and Leutenegger et al. (\cite{Leut06}). Because of photoexcitation of the
metastable 1s2s $^3S_1$ state, there are effectively only two lines in each
triplet, resonance ($r$) and intercombination ($i$). Note however that the
models {\it do} correctly account for the radial dependence of the
forbidden-to-intercombination line ratio, as described in Leutenegger et
al. (\cite{Leut06}). This can be seen in the weak, broad forbidden lines ($f$).

The models shown have been chosen to fit the red wing of the intercombination
line in order to show that the model resonance line is obviously too
blueshifted for both complexes. The best-fit models, presented in Leutenegger
et al. (\cite{Leut07}), have strong residuals in both lines. In both cases,
the model clearly does not fit the data, and the $r$ and $i$ lines clearly
have different shapes. The observed difference in the profile shapes is an
empirical fact, and it is independent of modelling assumptions.

This is remarkable, because we expect these lines to have almost the same
profile (modulo the small difference caused by the changing
forbidden-to-intercombination line ratio at large radii). This is because both
of these lines originate from transitions in the {\it same ion}. 

Resonance scattering can symmetrize lines with high optical depths. The
difference in observed profile shapes suggests that this effect is responsible
for the shape of the resonance line, which has a high oscillator strength,
while it does not affect the intercombination line, which has a low oscillator
strength. 

\section{Profiles including resonance scattering}

The theory of resonance scattering in O star X-ray line profiles is discussed
in Ignace \& Gayley (\cite{IG02}) and Leutenegger et al. (\cite{Leut07}). 
In both these analyses, radiative transfer is considered in the Sobolev
approximation.

One of the main results of Sobolev theory is that the optical depth of a
strong line at a given point depends on the local line-of-sight velocity
gradient.  In a spherically symmetric radial outflow, the velocity
gradient in the radial direction is just the velocity gradient of the flow,
$dv/dr$. In the lateral direction, the line-of-sight velocity gradient is a
consequence of the spherical divergence of the wind, $v/r$. 

For a wind with a $\beta = 1$ velocity law, $v/r > dv/dr$ beyond two
stellar radii. Far out in the wind, the radial velocity gradient is
negligible. Photons emitted primarily from far out in the wind thus see a much
higher optical depth in the radial direction than in the lateral direction,
and they escape preferentially in the lateral direction. Since the projected
velocity is the product of the outflow velocity times the direction cosine,
profiles formed by photons emitted preferentially in the lateral direction
will be more symmetric than profiles resulting from isotropic emission. This
will be the case in typical X-ray line profiles, as long as the continuum
optical depth is high enough to obscure photons coming from the inner part of
the wind. 

\begin{figure}[H]
  \begin{center}
    \includegraphics[width=\columnwidth]{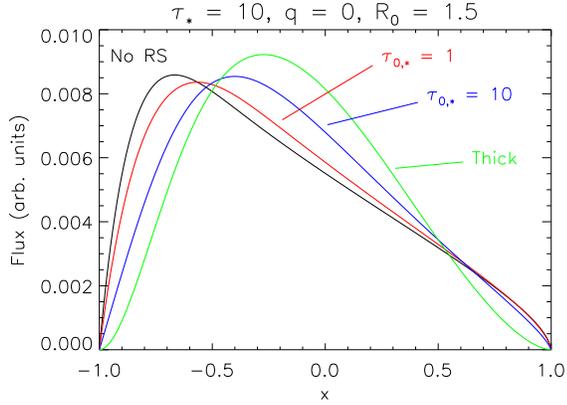}
    \caption{Comparison of different values of the characteristic line optical
      depth $\tau_{0,*}$.}
    \label{fig:t0s}
  \end{center}
\end{figure}

In Leutenegger et al. (\cite{Leut07}), we introduce a new free parameter,
$\tau_{0,*}$, which gives the characteristic Sobolev optical depth for a given
line. In Figure~\ref{fig:t0s} we plot model profiles with
different values of $\tau_{0,*}$. The models all have the same parameters,
other than the variation of the characteristic line optical depth. The effect of
resonance scattering is to significantly symmetrize a line profile, with
higher optical depths causing stronger symmetrization. For a more in depth
discussion of the derivation of this model and its parameters, see Leutenegger
et al. (\cite{Leut07}).

\section{Application of resonance scattering models to data}

\begin{figure}[H]
  \begin{center}
    \includegraphics[width=\columnwidth]{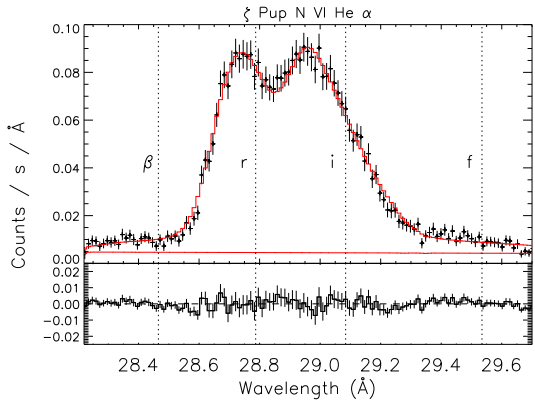}
    \caption{As Fig.~\ref{fig:nvi_nors}, but showing the best fit model
      including resonance scattering for \ionalt{N}{VI}.}
    \label{fig:nvi_b0}
  \end{center}
\end{figure}

\begin{figure}[H]
  \begin{center}
    \includegraphics[width=\columnwidth]{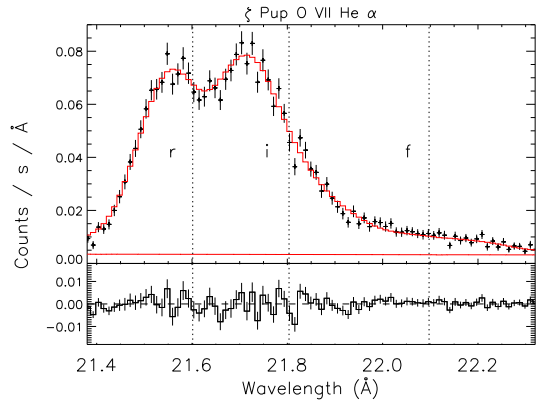}
    \caption{As Fig.~\ref{fig:nvi_b0}, but for \ionalt{O}{VII}.}
    \label{fig:ovii_b0}
  \end{center}
\end{figure}

In figures~\ref{fig:nvi_b0} and \ref{fig:ovii_b0} we show the best fit models
for \ionalt{N}{VI} and \ionalt{O}{VII} including the effects of resonance
scattering. These models obviously fit the data much better than the models
without resonance scattering. The fact that we observe the same effect in two
different ions suggests that this is not a spectroscopic artifact or anomaly.
The \ionalt{O}{VII} complex requires a moderate line optical depth, while the
\ionalt{N}{VI} complex requires a high line optical depth.

\section{Conclusions}

By comparing the shapes of line profiles from resonance and intercombination
lines originating from the same He-like ion in the X-ray spectrum of $\zeta$
Pup, we find that resonance scattering is important in the formation of the
resonance lines and causes them to be significantly more symmetric than the
intercombination lines.

This effect could be important for many lines in O stars with high mass-loss
rates, and any future analyses of line profiles should take it into account. 
However, we expect that this effect will be more marginal for K-shell lines of
Ne and higher Z elements and L-shell lines of Fe due to their lower elemental
abundances. 

Accounting for this effect in modelling will allow a partial reconciliation
with published mass-loss rates based on density-squared diagnostics, but will
likely still require reductions in O star mass-loss rates of a factor of a few. 
We estimate that our measured values of the continuum optical depth $\tau_*$
are consistent with a mass-loss rate of $\sim 1.5 \times 10^{-6}\, \mathrm{
  M_\odot\, yr^{-1}}$, which is a factor of a few greater than the mass-loss
rate suggested by modeling profiles without including the effect of resonance
scattering. Furthermore, we also estimate that our values of $\tau_*$ are
consistent with those inferred from the fits to the 15.014 \AA\ line of
\ionalt{Fe}{XVII} and the Ly$\alpha$ line of \ionalt{Ne}{X} in the {\it
  Chandra} HETG spectrum of $\zeta$ Pup presented by Cohen et al. in these
proceedings.  

\end{multicols}